\documentclass[conference, 10pt, a4paper]{IEEEtran}
\IEEEoverridecommandlockouts

\usepackage{microtype}
\usepackage{hyperref}
\usepackage[english]{babel}
\usepackage[T1]{fontenc}
\usepackage[utf8]{inputenc}
\usepackage{graphicx}
\usepackage{booktabs}
\usepackage[maxbibnames=9, minbibnames=3, backend=biber, doi=false, url=false, isbn=false]{biblatex}
\bibliography{literature}

\AtEveryBibitem{\clearfield{pages}}

\begin{document}
\title{Look Mum, no VM Exits! (Almost)}
\author{
\IEEEauthorblockN{
	Ralf Ramsauer\IEEEauthorrefmark{1},
	Jan Kiszka\IEEEauthorrefmark{2},
	Daniel Lohmann\IEEEauthorrefmark{3} and
	Wolfgang Mauerer\IEEEauthorrefmark{1}\IEEEauthorrefmark{2}}
\IEEEauthorblockA{
	\IEEEauthorrefmark{1}Technical University of Applied Sciences Regensburg\\
	\IEEEauthorrefmark{2}Siemens AG, Corporate Technology, Munich\\
	\IEEEauthorrefmark{3}University of Hanover\\
	ralf.ramsauer@othr.de,
	jan.kiszka@siemens.com,
	lohmann@sra.uni-hannover.de,
	wolfgang.mauerer@othr.de
\thanks{This work was partly supprted by the German Research Council (DFG) under grant no. LO 1719/3-1}
}}
\maketitle
\begin{abstract}
  Multi-core CPUs are a standard component in many modern embedded systems.
  Their virtualisation extensions enable the isolation of services, and gain
  popularity to implement mixed-criticality or otherwise split systems. We
  present Jailhouse, a Linux-based, OS-agnostic partitioning hypervisor that
  uses novel architectural approaches to combine Linux, a powerful
  general-purpose system, with strictly isolated special-purpose components.
  Our design goals favour simplicity over features, establish a minimal code
  base, and minimise hypervisor activity.

  Direct assignment of hardware to guests, together with a deferred
  initialisation scheme, offloads any complex hardware handling and
  bootstrapping issues from the hypervisor to the general purpose OS.  The
  hypervisor establishes isolated domains that directly access physical
  resources without the need for emulation or paravirtualisation.  This
  retains, with negligible system overhead, Linux's feature-richness in
  uncritical parts, while frugal safety and real-time critical workloads
  execute in isolated, safe domains.
\end{abstract}

\begin{figure*}[t]
	\includegraphics[width=1.0\textwidth]{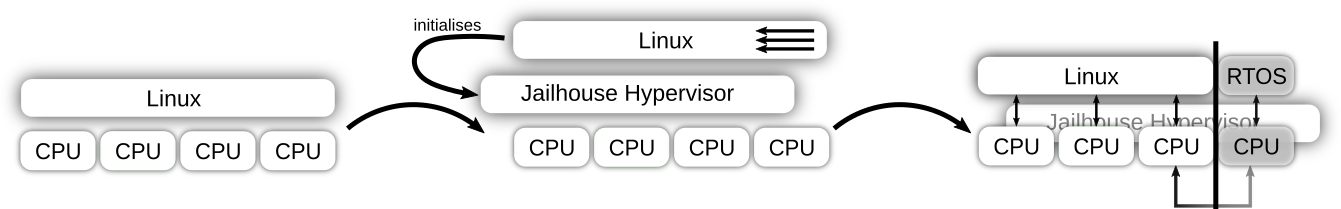}
	\caption{Activation sequence of the Jailhouse hypervisor.  After Linux
	has loaded and started the hypervisor, an additional real-time
	operating system is started in an isolated critical domain.}
	\label{fig:enabling}
\end{figure*}

\section{Introduction}
Despite the omnipresence of multi-core CPUs, manufacturers of safety critical
and uncritical products still tend to split components with different levels of
criticality to separate hardware units. In such traditional mixed criticality
environments, single logical control tasks are strongly bound to dedicated
physical control units.  Typical representatives of this architectural approach
range from automotive, where it is not uncommon that a single car contains
dozens to a hundred of different control units~\cite{broy:06:icse}, to
industrial Programmable Logic Controllers (PLCs), where \emph{critical} logical
control tasks are executed on a different physical computing platform than
\emph{uncritical} Human Machine Interfaces (HMIs).  Consolidating such systems
to single hardware units is an architectural trend~\cite{broy:06:icse} that
does not only improve the maintainability of substantial and growing amount of
software, but also reduces the overall hardware costs.

CPU virtualisation extensions can be exploited to create execution environments
that ease the port of existing legacy payload applications into strictly
isolated execution domains that cannot interfere with each other in an
unacceptable way.  Our approach relies on these widely available techniques to
statically partition hardware while reducing architecture dependencies.

The Linux-based hypervisor Jailhouse, initially developed by one of the authors
(JK) and subsequently refined as open-source software,\footnote{Available at
\url{https://github.com/siemens/jailhouse} under GPLv2.} is at the core of our
architecture.  It transforms symmetric multiprocessing (SMP) systems into
asymmetric multiprocessing (AMP) systems by inserting ``virtual barriers'' to
the system and the I/O bus.  From a hardware point of view, the system bus is
still shared, while software is \emph{jailed} in \emph{cells} from where the
guest software, so-called \emph{inmates}, can only reach a subset of physical
hardware.

Jailhouse is enabled by a kernel module from within a fully booted Linux
system, see Figure~\ref{fig:enabling}.  It takes control over all hardware
resources, reassigns them back to Linux according to a configuration of the
system, and lifts Linux into the state of a virtual machine (VM).  The
hypervisor core of Jailhouse acts as Virtual Machine Monitor (VMM).  This
scheme does not fit into the traditional classification of
hypervisors~\cite{goldberg73} -- it can be seen as a mixture of Type-1 and
Type-2 hypervisors: It runs on raw hardware like a bare-metal hypervisor
without an underlying system level, but still cannot operate without Linux as a
system aide to provide initialised hardware.  Linux is used as bootloader, but
not for operation.  Unlike other real-time partitioning approaches (e.g.,
PikeOS~\cite{pikeos}) that aim to manage hardware resources and may forbid
direct access by guest systems, Jailhouse \emph{only} supports direct hardware
access.  Instead of using complex and time-consuming
(para-)virtualisation~\cite{xen} schemes to emulate device drivers and share
physical hardware resources, Jailhouse follows an exokernel-like
approach~\cite{engler:95:sosp} in that it only provides isolation (by
exploiting virtualisation extensions) but intentionally neither provides a
scheduler nor virtual CPUs.  Only (few) resources that can, depending on the
hardware support, not yet be partitioned in that way are virtualised in
software.

For cost effectiveness, many industrial applications cannot give up on the
capabilities and feature-richness of Linux in their systems, yet they face
increasing demands to simultaneously cope with safety or other certification
requirements that are difficult to achieve with Linux.  Our architectural
approach fulfils these needs.  However, we consider it also an ideal framework
to ease the integration of state-of-the art research or experimental systems
that solve a specific problem in a novel way with industry-grade solutions
based on Linux.

In this paper, we present:
\begin{itemize}
	\item The architecture of Jailhouse, a fully functioning,
		non-scheduling, real-time, statically partitioning, and
		open-source hypervisor running on several architectures. 
	\item The implementation of a non-trivial real-world mixed-criticality
		application running as a Jailhouse guest.
	\item Advantages of \emph{deferred} hypervisor activation.
	\item A quintessential microbenchmark of the interrupt system on an
		Nvidia Jetson TK1
\end{itemize}

\section{Related Work}
Embedded virtualisation substantially differs from common enterprise, desktop
or mainframe virtualisation~\cite{heiser08-role}, where the technology has
its roots.  Many segments consider the consolidation of services as major
motivation.  While hypervisors are often optimised for high throughput and
optimal performance in the desktop and enterprise segment, virtualisation
solutions for real-time constrained embedded systems especially target low
latencies, deterministic computation cycles and maintaining real-time
capabilities~\cite{heiser08-role, pikeos, masmano2005, steinberg:10:eurosys,
xi11-xen}.

Nevertheless, many embedded hypervisors adopt established practices from
\emph{classical} virtualisation: overcommitting of hardware,
paravirtualisation~\cite{xen} or emulation of devices, and guest scheduling.

Crespo et al. present the XtratuM~\cite{xtratum} embedded hypervisor. Their
approach focuses on design constraints given by avionic guidelines and
specifications.  With memory management, clock and timer management, interrupt
management, a feature-rich hypercall interface and an own scheduler, XtratuM is
a full-fledged hypervisor.

The PikeOS~\cite{pikeos} real-time separation kernel approach allows for
executing different guest environments or native tasks.  For running guest
operating systems, PikeOS uses paravirtualisation and hardware-assisted
virtualisation, but also allows direct I/O access.  To payload applications,
PikeOS incorporates a combination of time- and priority driven scheduling, and
use best effort scheduling for uncritical tasks.

To implement temporal and spatial isolation, hypervisors do not always require
the availability of all virtualisation extensions.  Pinto et
al.~\cite{pinto2014} show an interesting approach by exploiting the ARM
TrustZone technology to run a real-time operating system in parallel to Linux
on a single CPU.  Their approach maintains real-time capabilities by using fast
interrupts (FIQs) only for real-time critical devices.  In contrast to regular
IRQs, those interrupts arrive directly in the secure world, where the real-time
operating system and the hypervisor execute.  Normal interrupts arrive in the
non-secure world, which is isolated from the secure world.  This approach only
isolates the non-secure from the secure world, and not vice versa.
Additionally, the TrustZone approach only allows for the creation of two
domains.

Quest-V~\cite{li:14:vee} is an advancement of the Quest operating system and
similar to Jailhouse in several respects.  It aims for static hardware
partitioning with minimum monitor activity.  In contrast to Quest-V, Jailhouse
is a VMM only, and does not implement any device drivers which drastically
minimises its code base.  Quest-V relies on paravirtualisation schemes to boot
Linux kernel as guest.

Jailhouse, in contrast to all those systems, starts with Linux (and exploits
its capabilities to initialize most of the hardware) and then uses deferred (or
late) hypervisor activation \cite{rutkowska06-bluepill} to partition the
hardware underneath the already running Linux.  \unskip\kern-2pt \footnote{To
the best of our knowledge, Rutkowska~\cite{rutkowska06-bluepill} was the first
who used this technique to inject undetectable malware (i.e., a thin
hypervisor) into computer systems.}

\section{Static Hardware Partitioning}
\subsection{Jailhouse Philosophy}
As is shown in Figure~\ref{fig:enabling}, activating the Jailhouse VMM is done
with the assistance of a Linux kernel module containing the hypervisor (HV).
After the HV startup code is executed by each CPU, Linux continues to run as a
virtual machine and \emph{guest} of Jailhouse, the so-called \textit{root
cell}.

The myriad of existing different hardware makes it hard or even impossible for
research groups with limited resources to support them in their systems.
Linux, on the contrary, is an extremely powerful operating system concerning
hardware support.  Jailhouse takes this advantage and hijacks Linux.  The
untypical \emph{deferred} activation procedure of the VMM has the considerable
practical advantage that the majority of hardware initialisation is fully
offloaded to Linux, and Jailhouse can entirely concentrate on managing
virtualisation extensions.  Similar to the exo-kernel~\cite{engler:95:sosp}
approach, Jailhouse is an \emph{exo-hypervisor}, with the difference that the
skeleton is modeled by the corpus, and not vice versa.  The direct assignment
of hardware devices allows Linux for continuing executing as before. Unlike
other partitioning approaches (for instance,~\cite{li:14:vee}), Jailhouse does
not require any specific device drivers except for minimalist, optional debug
helpers.

Jailhouse assumes that physical hardware resources do not need to be shared
across guests.  To create additional domains (called \textit{non-root cells}),
Jailhouse removes hardware resources (e.g., CPU(s), memory, PCI or MMIO
devices) from Linux and reassigns them to the new domain.  Therefore Linux
releases the hardware if it has previously been in use.  This includes physical
CPUs: the configuration of a partition consists at least of one CPU and a
certain amount of memory that is preloaded by the root cell with a secondary
operating system or a bare-metal application.

Linux offlines selected CPUs and calls the hypervisor to create a new cell by
providing a \emph{cell configuration} that describes the assigned resources.
Other resources, like PCI devices, memory-mapped devices or I/O ports, can be
exclusively reassigned to the new guest as well.  The hypervisor prevents
subsequent access to those resources from any other domain, which prohibits
inadvertent modifications.  Non-root cells can dynamically be created,
destroyed (i.e., resources are assigned back to the root cell) or relaunched.

Virtualisation extensions (See Ref.~\cite{arm, svm, vtx} for the four major
architectures ARMv7 with Virtualization Extensions (VE), ARMv8, Intel 64-bit
x86 with VT-x and VT-d support, and amd64 with SVM support) guarantee spatial
isolation: any access violation, for instance illegal access across partitions,
\textit{traps}~\cite{popekformal} the hypervisor, which eventually stops
execution.  Certain instructions executed by guests cause traps and must be
handled by the hypervisor.

Since Jailhouse only remaps and reassigns resources, the ideal design
conception is that -- besides management -- it does not need to be active after
setting up and starting all guests, and only intercepts in case of access
violations: \enquote{Look Mum, no VM Exits!} However, hardware is not (yet)
perfectly suited for this approach, so on current hardware, the following
circumstances still require intervention by the VMM:
\begin{itemize}
	\item Interrupt reinjection (depending on the architecture,
		interrupts may not directly arrive at guests)
	\item Interception of non-virtualisable hardware resources\\
		(e.g., parts of the Generic Interrupt Controller (GIC) on ARM)
	\item Access of platform specifics (e.g., accessing Control Coprocessor
		CP15 or Power State Control Interface (PSCI) on ARM)
	\item Emulation of certain instructions (e.g., \texttt{cpuid} on x86)
\end{itemize}
The following traps are unavoidable, and not contrary to our concept, as they
only occur in case of \emph{jailbreak} or cell management:
\begin{itemize}
	\item Access violations (memory, I/O ports)
	\item Cell management (e.g., creating, starting, stopping or destroying
		cells)
\end{itemize}
These interceptions introduce overhead and latencies -- virtualisation, of
course, comes at a cost~\cite{drepper:08:acmqueue}.  In
section~\ref{sec:evaluation} we exemplarily present the evaluation of one
fundamental microbenchmark, the additional latency of interrupts.

Despite the strict segregation of resources across guests, Jailhouse still
allows cells to share share physical pages.  Besides enabling inter-cell
communication, the mechanism also allows for sharing memory-mapped I/O pages,
which, if desired, allows for accessing hardware resources from within multiple
domains.  Such concurrent access is, however, not arbitrated by Jailhouse and
needs to be addressed appropriately by the guests.

\begin{figure}[t]
\includegraphics[width=1.0\columnwidth]{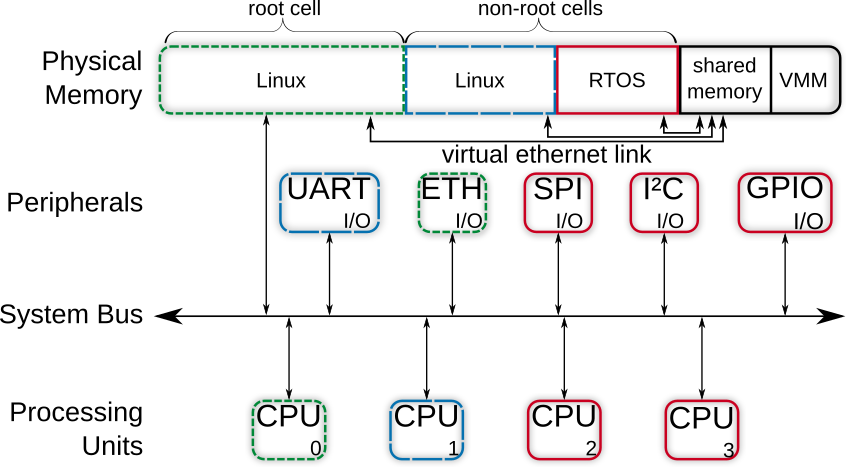}
\caption{Concept of ideal hardware partitioning: while the system bus is
	still shared, the Jailhouse hypervisor takes care that cells only
	access resources within their scope (colored).  Safe communication
	between domains is enabled by shared memory.}
\label{fig:arch}
\end{figure}

Figure~\ref{fig:arch} shows a possible partitioned system layout for three
cells: the Linux root cell (green), an additional Linux non-root cell (blue)
and a minimalist real-time operating system (red).  Communication between
cells is realised by shared memory regions, together with a signalling
interface.  This minimalist design requires no additional device driver logic
in the hypervisor.  Depending on the hardware support, it is implemented based
on a virtual PCI device through Message-Signaled Interrupts (MSI-X) or legacy
interrupts.  A guest may use this device to implement a virtual ethernet device
on top of it.  On systems without PCI support, Jailhouse emulates a generic and
simple PCI host controller.  We chose emulation in this case, as PCI provides
a configuration space: The PCI device identifies itself and its capabilities.
This enables, if supported, automatic configuration in guests, and the virtual
PCI host controller results in only six lines of code and does not increase the
overall code
size\footnote{\url{https://github.com/siemens/jailhouse/commit/7b9f373dcfc14a4951928c43ded9c02b9f1ac02c}}.

\subsection{Support}
The partitioning approach allows a safety-certified operating system or a
bare-metal application to run on a consolidated multi-core system in parallel
to Linux.  It is worth mentioning that despite the fact that Jailhouse supports
four different CPU architectures, which goes beyond what is provided by many
experimental or research systems, its minimalist approach results in only a
few thousands lines of code for the core parts. This simplifies certification
processes, but allows developers to concentrate on important issues without
spending time on providing a never complete number of device drivers that are
required to make the system usable in realistic environments. The simplicity of
the core is a good basis for a formal verification of the hypervisor, similar
to the formal verification of related system
software~\cite{klein09-verification}.

Jailhouse comes with its own \emph{inmate} library that allows for running
minimalistic demo applications.  Several operating systems apart from Linux are
already available as Jailhouse guests (L4 Fiasco.OC on
x86~\cite{baryshnikov}, FreeRTOS on ARM, Erika Enterprise RTOS v3 on ARM64).  We
have successfully ported the RTEMS real-time operating system for the ARM
architecture with very limited effort -- modifications are mostly required for
platform specific board support.  The simplicity of porting systems suggests an
opportunity to expose feature-incomplete research approaches to realistic
industrial use-cases by combining them with an industrial grade base.

\subsection{Practicability}
To demonstrate the suitability of our approach especially for practical use, we
implemented a (mixed-criticality) multi-copter control system.  The
requirements on such platforms are comparable to many common industrial
appliances: The flight stack, a safety and real-time critical part of the
system with high reliability requirements, is responsible for balancing and
navigating the aircraft.  Sensor values must be sampled at high data rates,
processed, and eventually be used to control rotors.  For a safe and reliable
mission, the control loop must respond deterministically.  System crashes may
result in real crashes with severe consequences.

The flight stack runs in a Jailhouse cell, while uncritical tasks, for example
WiFi communication with a ground station or camera tracking, can easily be
implemented in an uncritical part thanks to the available Linux software
ecosystem.  Critical hardware components, e.g., SPI, I$^2$C or GPIO devices,
are assigned to the critical cell.  Our hardware platform is an Nvidia Jetson
TK1 with a quad-core Cortex-A15 ARMv7 CPU, connected to a sensor board that
provides accelerometers, GPS, compasses and gyroscopes.  Two cores are assigned
to the uncritical part, and two cores to the critical one.

The critical domain executes a second stripped-down Linux operating system with
the Preempt\_RT real-time kernel extension.  Ardupilot provides flight control,
and does not require modifications besides board support.  This underlines that
existing applications can be deployed in a Jailhouse setup with little effort,
and that it is suitable for implementing real-time safety critical systems
mostly based on existing components.  Nonetheless, we needed to solve various
issues that do not arise on a purely conceptual level or with systems tailored
for very specific hardware, but endanger assumptions made in our and similar
approaches.

\section{Evaluation}
\label{sec:evaluation}
As mentioned before, the aim of Jailhouse is to minimise the activity of the
hypervisor.  Though this would be possible in theory, the sole existence of a
hypervisor introduces additional latencies~\cite{drepper:08:acmqueue} that do
not exist without a VMM.  For example, shadow page tables may introduce
additional memory access latencies.

To evaluate and determine the (real-time) performance of the hypervisor,
several environmental conditions must be considered.  It is hard or even
impossible to quantify the hypervisor overhead with one single scalar.  This
results in a set of microbenchmarks.

For all benchmarks, single-shot measurements do not allow to draw any
conclusions on the behaviour of the system.  Microbenchmarks should be repeated
under certain environmental conditions, such as the actual existence of a
hypervisor, and the particular frequency of a certain measurement together with
the utilisation of other guests.

Due to the limited size of the paper, we will exemplarily present the
measurement of the interrupt latency in detail, and describe other considerable
measurements.

It is important to remark that such benchmarks do not measure the overhead of
the hypervisor, but the overhead of the hypervisor when running on a
\emph{specific} hardware platform.  Still, those measurements allow to derive a
trend of the performance of the hypervisor.

\paragraph{Hypercalls} One typical benchmark for hypervisors is the cost of
hypercalls.  In case of Jailhouse, hypercalls do not need to be considered, as
they are only used for cell management purposes, and never occur in hot paths.

\paragraph{Shared System Bus} Different guests asynchronously access memory,
and memory or I/O access may be serialised by hardware.  Though starvation does
not occur on supported architectures,  heavy utilisation of memory or I/O
busses may lead to significant slow downs of guests.  While this problem is
well-known for SMP applications, its impact must be evaluated when
asynchronously executing multiple payloads that were designed for single-core
platforms.

\paragraph{Architecture-dependent Traps} Because of architectural limitations,
Jailhouse needs to emulate devices that are essential for a hardware platform
and that cannot be virtualised in hardware (e.g., the interrupt distributor as
part of the generic interrupt controller on ARM architectures).  Depending on
the utilisation of those devices, the impact of the hypervisor must be
analysed.

\paragraph{Interrupt Latency}
Jailhouse supports two versions of ARM's Generic Interrupt Controller, GICv2
and GICv3~\cite{gicv2, gicv4}.  Both implementations share the same
architectural limitation: Interrupts do not directly arrive at the guest.  They
arrive at the hypervisor, and are then reinjected as virtual IRQs to the guest.
This leads to an overhead in the hypervisor, as it must redirect the interrupt
to the appropriate guest, followed by a switch of the privilege level.

Our automated measurement setup consists of an Nvidia Jetson TK1 (quad-core
Cortex-A15 @2.32GHz) as target platform, and an Arduino Uno for performing the
actual measurement.

To measure this latency, we compare the \emph{bare-metal latency} (i.e., the
minimum latency without hypervisor) with the latency when the hypervisor is
present.  The Arduino periodically toggles a GPIO pin on the target board which
causes an interrupt.  The only task of the non-root cell is to answer as soon
as possible to the interrupt by toggling another GPIO.  Therefore, we
implemented a minimalistic guest that uses Jailhouse's own inmate library.
To minimise code size for the response to make it as fast as possible, the
instructions for toggling the GPIO are directly written in assembler in the
interrupt vector table.  The measurement without hypervisor represents the
\emph{bare minimum} latency achievable by the selected hardware platform.
Latency difference with and without hypervisor presence measures the delay that
is introduced when the hypervisor and other guests asynchronously access the
system bus.  The Capture Compare Unit of the Uno ensures a precise measurement
at a resolution of 62.5ns.  To validate measurements, we verified sample
measurements with the latency manually measured by an oscilloscope.

We repeat the measurement under several conditions (e.g., load is placed on
other guests to measure the influence on the shared system bus) and present the
arithmetic mean as well as the standard deviation and the maximum latency.
Every measurement runs for four hours, and was repeated with an interrupt
frequency of 10Hz and 50Hz to determine the role of the frequency of the
measurement.  The \emph{stress} parameter in Table~\ref{table:int} describes if
other guests are put under CPU, I/O or memory load with the stress-ng
benchmark.

Results can be found in Table~\ref{table:int}.  The first two lines show the
minimum interrupt latency of the measurement without the existence of the
hypervisor.  The difference to other measurements denotes the overhead that is
introduces by the hypervisor.

The latency that is introduced by the hypervisor does not significantly depend
on the interrupt frequency, but on the utilisation of neighbouring guests.
This effect is caused by the shared system bus: The hypervisor wants to access
memory that is required for dispatching the interrupt, while other guests
asynchronously access the same bus.

On average, interrupt latency amounts to $\approx810$ns, with narrow deviation.
Still, outliers lead to latencies of almost 5$\mu$s.  Compared to the cycle
times of typical industrial communication bus systems, the maximum delay is
acceptable for many applications.

\begin{table}
	\centering
	\caption{Interrupt Latency on an Nvidia Jetson TK1 (in $\mu$s)}
	\label{table:int}
	\begin{tabular}{rrrrrr}
		\toprule
		VMM	& Freq	& Stress & $\mu$	& $\sigma$	& Max\\
		\midrule
		off	& 10Hz & no  & 0.45 & 0.02 & 0.50\\
		off	& 50Hz & no  & 0.45 & 0.02 & 0.50\\
		on	& 10Hz & no  & 1.26 & 0.07 & 2.81\\
		on	& 50Hz & no  & 1.25 & 0.04 & 2.94\\
		on	& 10Hz & yes & 1.36 & 0.34 & 5.56\\
		on	& 50Hz & yes & 1.35 & 0.34 & 5.38\\
		\bottomrule
	\end{tabular}
\end{table}

\section{Discussion}
The minimalist design approach of Jailhouse results in a manageable amount of
source lines of code (SLOC).  This is a crucial factor for both, formal
verification from an academic point of view and system certification from an
industrial point of view.\footnote{We are aware of the problem that a
substantial chain of software besides the Linux kernel (e.g., UEFI firmware
code, bootloaders etc.) is required for the boot process, and needs to be
considered in such certifications to some extent. There are various possible
approaches to address these issues that will be discussed in future work.}

Jailhouse, in total, consists of almost 30k SLOC for four different
architectures.  This includes the hypervisor core, example code, kernel driver,
and userland tools and utilities.  Substantial parts of the code are
architecture-independent.  The common critical hypervisor core code that is
shared across all architectures amounts to less than 3.4k SLOC.  Architecture
dependent code amounts to $\approx$7.4k SLOC for x86 and implements both, Intel
and AMD, and $\approx$5.4k SLOC for ARM (both, ARMv7 and ARMv8).  Exemplarily,
the whole hypervisor core for ARMv7 sums up to $\approx$7.4k SLOC.

Many research systems are developed from scratch and spend tremendous effort on
re-implementing existing device drivers.  But still, missing device support is
a major obstacle for their practicability.  More than half of Quest-V's source
lines of code ($\approx$70k SLOC of 140k SLOC) implement device drivers.  With
almost 27k SLOC, XtratuM is more lightweight than Quest-V and only implements
basic drivers for debug output.  Still, the publicly available versions of
Quest-V and XtratuM currently only support the x86 architecture.

Jailhouse does intentionally not follow classical virtualisation approaches,
but its design does not generally eliminate the use of those techniques.  This
opens the possibility to exploit Jailhouse as an experimental systems platform
that allows for keeping focus on the actual problem instead of re-implementing
fundamentals from scratch.  Jailhouse is an ideal platform for investigating
hardware and software behaviour under AMP workloads.  Furthermore, it provides
a convenient and comfortable environment for executing digital signal
processing (DSP)-like workloads on \emph{raw} hardware.

Modern multi-core systems already provide enough physical CPUs to make
scheduling in hypervisors unnecessary for many real-world embedded use cases.
In fact, numerous essential requirements on real-time embedded
hypervisors~\cite{xtratum}, such as real-time scheduling policies, efficient
context switches, or deterministic hypervisor calls, do not even need to be
addressed in a partitioned setup.  Those requirements actually reflect
well-known issues of operating systems and should not be duplicated in
hypervisor space for embedded systems with real-time requirements.  As
Jailhouse does not virtualise CPUs, overcommit hardware or schedule partitions,
there are no expensive partition context switches or scheduling
issues~\cite{vestal07-preemptive} as they occur in other real-time
hypervisors~\cite{pikeos, masmano2005, pinto2014, xi11-xen}.  Hypercalls are
only used for management purposes and not for moderating access to shared
hardware.

Depending on the interrupt system and the architecture, interrupts might arrive
at the hypervisor.  On such platforms, the interrupt reinjection to guests is a
frequent job of the hypervisor that introduces unintended additional interrupt
latencies.  This issue is already solved for 64-bit x86 architectures that
support \emph{Interrupt Remapping} and will be solved in future ARM
architectures that implement the GICv4~\cite{gicv4} specification, which is
beneficial to our final goal, to end up in no VM exits.

Nevertheless, there are unavoidable traps that are caused by hardware design.
On current ARM architectures, the interrupt distributor must be virtualised.
Varanasi and Heiser~\cite{heiser11-ve} assume that this is not expected to
cause performance issues.  During the implementation of our demonstration
platform we contrariwise observed that Linux kernels with the Preempt\_RT
real-time patch make heavy use of the interrupt distributor which causes high
activity of the hypervisor.  Such issues should be addressed by proper hardware
design in order to be able to execute unmodified guests,

\section{Conclusion and Future Work}
The static partitioning hypervisor technique is a promising approach for
embedded real-time virtualisation, as their ultimate goal to minimise the
interaction with guests defers all issues that are introduced by typical
paravirtualisation approaches back to the operating systems of the guests,
where they already existed before.  The driverless approach tries to fill the
gap between academic research systems and industrial practicability.

In comparison to paravirtualisation techniques, direct hardware assignment to
guests allows for running unmodified legacy payload applications with no active
hypervisor overhead.  The minimalist hypervisor core simplifies certification
efforts.  By executing standard operating systems as guests, we also minimised
the effort that is required for porting existing legacy payload applications.
By implementing a complex demonstration platform, we successfully showed
the practicability of hardware partitioning.

While standard virtualisation extensions provided by current hardware seem to
suffice for a straight forward implementation of our and many other approaches,
real hardware presents a number of limitations that can completely undermine
the advantages and guarantees of partitioning and virtualisation-based
approaches.  Our future work will address arising issues and  concentrate on
evaluating the performance of the hypervisor.

\printbibliography
\end{document}